\documentclass[aps,prl,reprint,showpacs,floatfix]{revtex4-1}
\usepackage{graphicx, amsmath}
\usepackage[caption=false]{subfig}
\usepackage{pdfpages}
\begin{document}

%-------------------------------------------------------------------------------
% Shortcut Commands
%-------------------------------------------------------------------------------

\newcommand{\comb}[2]{{\begin{pmatrix} #1 \\ #2 \end{pmatrix}}}
\newcommand{\braket}[2]{{\left\langle #1 \middle| #2 \right\rangle}}
\newcommand{\bra}[1]{{\left\langle #1 \right|}}
\newcommand{\ket}[1]{{\left| #1 \right\rangle}}
\newcommand{\ketbra}[2]{{\left| #1 \middle\rangle \middle \langle #2 \right|}}

%-------------------------------------------------------------------------------
% Front Matter
%-------------------------------------------------------------------------------

\title{Connectivity is a Poor Indicator of Fast Quantum Search}

\author{David A. Meyer}
	\affiliation{Department of Mathematics, University of California, San Diego, La Jolla, CA 92093-0112}
	\email{dmeyer@math.ucsd.edu}
\author{Thomas G. Wong}
	\affiliation{Department of Physics, University of California, San Diego, La Jolla, CA 92093-0354}
	\altaffiliation{Currently at the University of Latvia}
	\email{twong@lu.lv}

\begin{abstract}
	A randomly walking quantum particle evolving by Schr\"odinger's equation searches on $d$-dimensional cubic lattices in $O(\sqrt{N})$ time when $d \ge 5$, and with progressively slower runtime as $d$ decreases. This suggests that graph connectivity (including vertex, edge, algebraic, and normalized algebraic connectivities) is an indicator of fast quantum search, a belief supported by fast quantum search on complete graphs, strongly regular graphs, and hypercubes, all of which are highly connected. In this paper, we show this intuition to be false by giving two examples of graphs for which the opposite holds true:~one with low connectivity but fast search, and one with high connectivity but slow search. The second example is a novel two-stage quantum walk algorithm in which the walking rate must be adjusted to yield high search probability.
\end{abstract}


\maketitle

%-------------------------------------------------------------------------------
% Main Matter
%-------------------------------------------------------------------------------

\textit{Introduction.}---Despite ten years elapsing since the introduction of continuous-time quantum walk algorithms that search on graphs \cite{CG2004}, there is still no comprehensive theory as to which graphs support fast quantum search. Nevertheless, much work has been done to further our understanding. For example, we recently showed that global symmetry is unnecessary for fast quantum search~\cite{JMW2014}.

\begin{figure}
\begin{center}
	\subfloat[]{
		\includegraphics{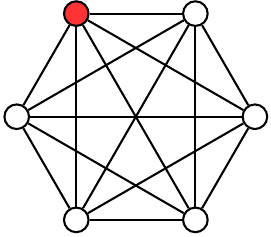}
	}
	\subfloat[]{
		\includegraphics{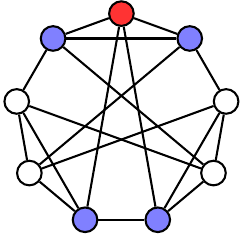}
	}
	\subfloat[]{
		\includegraphics{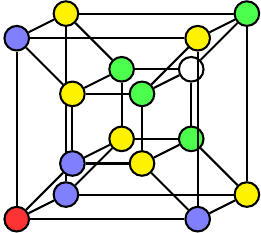}
	}
	\caption{\label{fig:graphs} (a) The complete graph with 6 vertices. (b) A strongly regular graph (Paley graph) with parameters (9,4,1,2). (c) The 4-dimensional hypercube. Without loss of generality, a marked vertex is colored red, and identically evolving vertices are identically colored.}
\end{center}
\end{figure}

Regarding specific graphs, a randomly walking quantum particle evolving by Schr\"odinger's equation searches on the complete graph, strongly regular graphs, and the hypercube in optimal $\Theta(\sqrt{N})$ time, the first of which is precisely the continous-time analogue of Grover's algorithm \cite{Grover1996, FG1998, CG2004, JMW2014}. Examples of these graphs are shown in Fig.~\ref{fig:graphs}. Additionally, such a particle can search on $d$-dimensional cubic lattices in $\Theta(\sqrt{N})$ total time when $d \ge 5$, and with progressively slower runtimes as $d$ decreases \cite{CG2004, Childs2003, Childs2012}, as shown in Table \ref{table:cubiclattices}.

\begin{figure}
\begin{center}
	\includegraphics{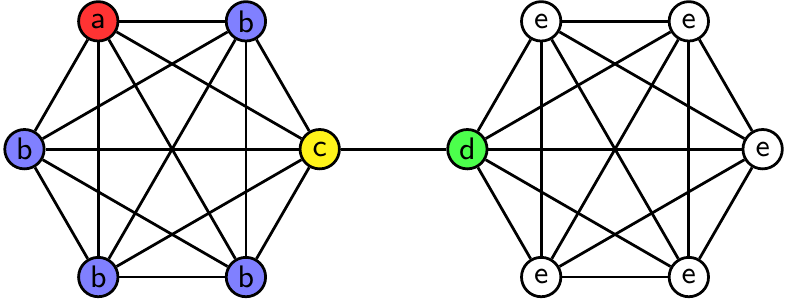}
	\caption{\label{fig:joined} A graph with 12 vertices constructed by joining two complete graphs of 6 vertices by a single edge.}
\end{center}
\end{figure}

\begin{figure}
\begin{center}
	\includegraphics{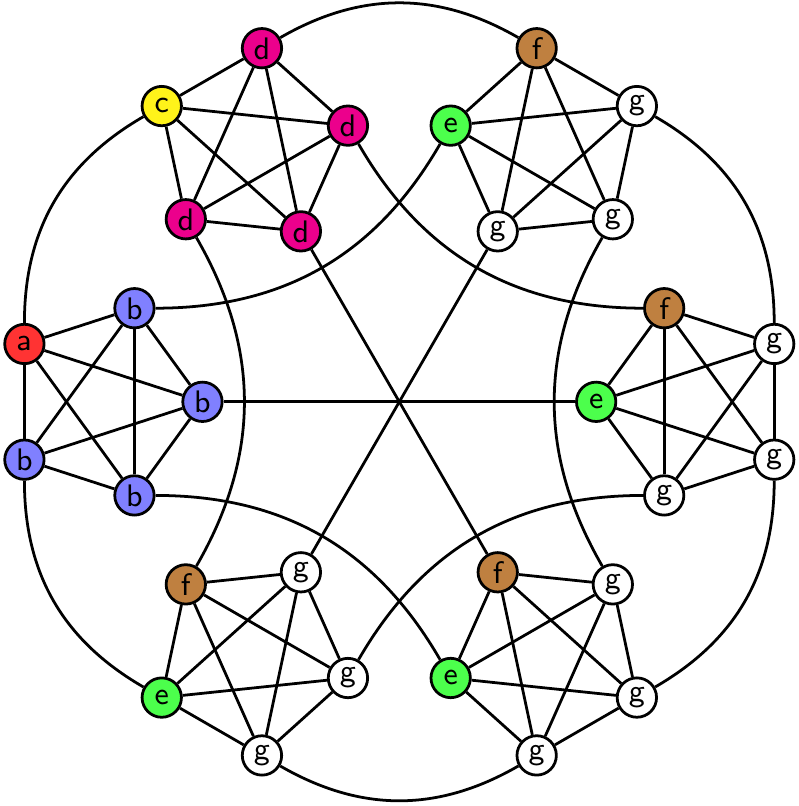}
	\caption{\label{fig:simplex} A 5-simplex with each vertex replaced with a complete graph of 5 vertices.}
\end{center}
\end{figure}

One might suspect that fast search occurs when graphs are highly connected, as suggested by \cite{CG2004}. In this paper, however, we show this intuition to be false by giving two examples of graphs for which the opposite holds true:~one with low connectivity but fast search, and one with high connectivity but slow search; they are shown in Figs.~\ref{fig:joined} and \ref{fig:simplex}, respectively. To do this, we first introduce four different ways to measure graph connectivity. Then we detail how a randomly walking quantum particle searches on a graph. Finally, we determine the runtimes of our two examples.

\begin{table}
\caption{\label{table:cubiclattices}Scalings of single runtimes and success probabilities for search on a $d$-dimensional cubic lattices by quantum random walk, and the total runtimes with classical repetitions.}
\begin{ruledtabular}
\begin{tabular}{cccc}
	$d$ & Single Runtime & Success Prob & Total Runtime \\
	\colrule
	$\ge 5$ & $N^{1/2}$ & $1$ & $N^{1/2}$ \\
	$4$ & $\sqrt{N \log N}$ & $1 / \log N$ & $\sqrt{N} \log^{3/2} N$ \\
	$3$ & $N^{2/3}$ & $1/N^{1/3}$ & $N$ \\
	$2$ & $N / \log N$ & $(\log^2 N)/ N$ & $N^2 / \log^3 N$ \\
\end{tabular}
\end{ruledtabular}
\end{table}

%-------------------------------------------------------------------------------
% Section.
%-------------------------------------------------------------------------------

\textit{Measures of Connectivity}.---The two most common ways to measure connectivity are \textit{vertex connectivity} and \textit{edge connectivity}, which are how many vertices or edges must be removed to make a graph disconnected. For example, Fig.~\ref{fig:joined} has vertex and edge connectivities of 1 because removing the yellow or green vertex disconnects the graph, and so does removing the edge between them. Note that vertex connectivity is upper bounded by the edge connectivity, and both are upper bounded by the minimum degree of the graph. For the graphs in this paper, the vertex and edge connectivities are equal.

Connectivity can also be measured using eigenvalues. The \textit{algebraic connectivity} of a graph is the second-smallest eigenvalue $\lambda_1$ of its graph Laplacian $L = D - A$, where $D_{jj} = \deg(j)$ is a diagonal matrix with the degree of each vertex, and $A_{ij} = 1$ if $i$ and $j$ are adjacent and $0$ otherwise is the adjacency matrix \cite{Fiedler1973}. Chosen this way, $L$ is positive semi-definite. Its smallest eigenvalue $\lambda_0$ is $0$, and it corresponds to the equilibrium state of diffusion. Since our graphs are connected, $\lambda_1$ is positive and quantifies how well diffusion occurs on the graph.
	
This can be improved by dividing $L_{ij}$ by $\sqrt{\deg(i) \deg(j)}$ so that the diagonal terms become $1$ and the off-diagonal terms become $-1 / \sqrt{\deg(i) \deg(j)}$ when $i$ and $j$ are adjacent and $0$ otherwise. The eigenvalues of this ``normalized'' Laplacian $\mathcal{L}$ are no longer dependent on the number of vertices, and we take the second-smallest eigenvalue to be the \textit{normalized algebraic connectivity} \cite{Chung1997}. Note if the graph is $k$-regular, then the normalized Laplacian is related to the adjacency matrix and standard Laplacian by $\mathcal{L} = I - A/k = L/k$.

For the graphs we have introduced, all four of these connectivities are shown in Table \ref{table:connectivities}; see \footnote{See Supplemental Material at [URL] for references and derivations of graph connectivities, and for proofs of runtimes for search on joined complete graphs and the simplex of complete graphs using degenerate perturbation theory.} for their references and derivations. With these in place, let us introduce the quantum search model and then find the runtimes of the examples (Figs.~\ref{fig:joined} and \ref{fig:simplex}), showing they are faster or slower, respectively, than their connectivities might otherwise lead us to believe.

\begin{table*}
\caption{\label{table:connectivities} The degrees and vertex, edge, algebraic, and normalized algebraic connectivities of various (nearly) regular graphs with $N$ vertices.}
\begin{ruledtabular}
\begin{tabular}{ccccc}
	Graph & Degree & Vertex/Edge & Algebraic & Normalized Algebraic \\
	\colrule
	Complete & $N-1$ & $N-1$ & $N$ & $N/(N-1) = \Theta(1)$ \\
	Strongly Regular (Type 1) & $(N-1)/2$ & $(N-1)/2$ & $(N - \sqrt{N})/2$ & $(N - \sqrt{N})/(N - 1) = \Theta(1)$ \\
	Strongly Regular (Latin Square) & $d(\sqrt{N}-1)$ & $d(\sqrt{N}-1)$ & $(d-1)\sqrt{N}$ & $(d-1)\sqrt{N} / [d(\sqrt{N}-1)] = \Theta(1)$ \\
	Hypercube & $\log_2 N$ & $\log_2 N$ & $2$ & $2 / \log_2 N = \Theta \left( 1 / \log N \right)$ \\
	$d$-dim Cubic & $2d$ & $2d$ & $\approx 4 \pi^2 / N^{2/d}$ & $\approx 2 \pi^2 / d N^{2/d} = \Theta( 1 / N^{2/d} )$ \\
	Joined Complete & $\approx N/2$ & $1$ & $\Theta(1)$ & $\Theta(1/N)$ \\
	Simplex Complete & $M = \Theta(\sqrt{N})$ & $M = \Theta(\sqrt{N})$ & $1$ & $1/M = \Theta(1/\sqrt{N})$
\end{tabular}
\end{ruledtabular}
\end{table*}

%-------------------------------------------------------------------------------
% Section.
%-------------------------------------------------------------------------------

\textit{Quantum Search on Graphs}.---The vertices of the graph correspond to computational basis states $\{ \ket{0}, \ket{1}, \dots, \ket{N-1} \}$ of an $N$-dimensional Hilbert space. The system $\ket{\psi(t)}$ begins in an equal superposition of all the vertices $\ket{s} = \sum_{i = 0}^{N-1} \ket{i} / \sqrt{N}$.
%\[ \ket{\psi(0)} = \ket{s} = \frac{1}{\sqrt{N}} \sum_{i = 0}^{N-1} \ket{i}. \]
Then it evolves by Schr\"odinger's equation with Hamiltonian $H = \gamma L - | a \rangle \langle a |$, where $\gamma$ is the jumping rate (\textit{i.e.}, amplitude per time), $L = D - A$ is the graph Laplacian, and $| a \rangle$ is the ``marked'' vertex we are searching for (\textit{i.e.}, the red vertex in Figs.~\ref{fig:graphs}, \ref{fig:joined}, and \ref{fig:simplex}). For a $k$-regular graph, $D = kI$, so we can drop it by rezeroing the energy. Then the search Hamiltonian is
\begin{equation}
	\label{eq:H}
	H = -\gamma A - | a \rangle \langle a |.
\end{equation}

\begin{figure}
\begin{center}
	\includegraphics{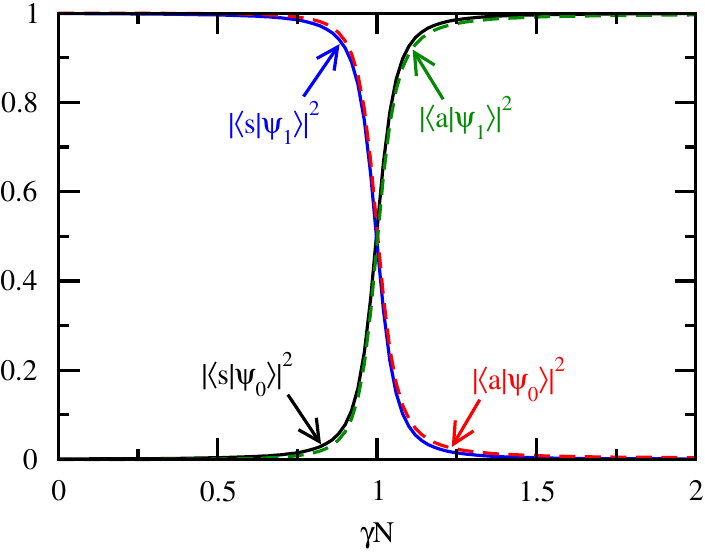}
	\caption{\label{fig:complete_overlap} Squared overlaps of $\ket{s}$ and $\ket{a}$ with eigenstates of $H$ for search on the complete graph with $N = 1024$.}
\end{center}
\end{figure}

On the complete graph (\textit{i.e.}, Grover's problem), the system evolves in a 2-dimensional subspace, and the squared overlaps of $\ket{s}$ and $\ket{a}$ with the eigenstates $\ket{\psi_{0,1}}$ of $H$ with  are shown in Fig.~\ref{fig:complete_overlap}. When $\gamma$ is away from its critical value of $\gamma_c = 1/N$, then the initial equal superposition $\ket{s}$ is approximately $\ket{\psi_0}$ or $\ket{\psi_1}$ for large $N$, so the system approximately evolves only by phase multiplication. When $\gamma = \gamma_c$, however, the eigenstates are
%On the complete graph (\textit{i.e.}, Grover's problem), the system evolves in a 2-dimensional subspace, and the squared overlaps of $\ket{s}$ and $\ket{a}$ with the eigenstates $\ket{\psi_{0,1}}$ of $H$ with  are shown in Fig.~\ref{fig:complete_overlap}. When $\gamma$ is away from its critical value of $\gamma_c = 1/N$, then the initial equal superposition $\ket{s}$ is approximately an eigenstate of $H$ for large $N$, so the system approximately evolves only by phase multiplication. When $\gamma = \gamma_c$, however, the eigenstates are
\begin{equation}
	\label{eq:complete_sa}
	\begin{aligned}
		&\ket{s} \propto \ket{\psi_0} + \ket{\psi_1} \\
		&\ket{a} \propto \ket{\psi_0} - \ket{\psi_1} 
	\end{aligned}
\end{equation}
with an energy gap of $\Delta E = 2/\sqrt{N}$ \cite{CG2004}. So the system evolves from $\ket{s}$ to $\ket{a}$ in time $\pi / \Delta E = \pi \sqrt{N} / 2 = \Theta(\sqrt{N})$ \cite{FG1998}. This can also be proved using degenerate perturbation theory \cite{JMW2014}, as we show rigorously for the next two examples in \footnotemark[1], but in this paper we use the same graphical explanation as above.

%-------------------------------------------------------------------------------
% Section.
%-------------------------------------------------------------------------------

\textit{Joined Complete Graphs}.---For the first example, two complete graphs with $N/2$ vertices are joined by a single edge. We mark a vertex away from this ``bridge'' so that it is one of $N - 2 = \Theta(N)$ possible vertices, as opposed to one of 2 vertices on the bridge, which would be a trivial search problem. In Fig.~\ref{fig:joined}, the marked vertex is colored red, and identically evolving vertices are the same color.

Intuitively, the bridge restricts probability from moving between the two complete graphs, so we are effectively searching on a single complete graph with $N/2$ vertices and total probability $1/2$. Thus the success probability should reach $1/2$ in time $\pi\sqrt{N/2}/2$, which is a total runtime of $\Theta(\sqrt{N})$ with the expected constant number of repetitions to boost the success probability near $1$. This is the same optimal runtime as the highly connected complete graph, strongly regular graphs, and hypercube, even though the vertex/edge and normalized algebraic connectivities suggest it should be slower. Note this example does not discredit algebraic connectivity since the hypercube is also $\Theta(1)$, but the second example will.

To prove this intuition, note from Fig.~\ref{fig:joined} that the system evolves in a 5-dimensional subspace, independent of $N$. Grouping identically-evolving vertices, we get an orthonormal basis for this subspace:
%\begin{align*}
%	&\ket{a} = \ket{\text{red}}, && \ket{d} = \ket{\text{green}}, \\
%	&\ket{b} = \frac{1}{\sqrt{N/2 - 2}} \sum_{i \in \text{blue}} \ket{i}, && \ket{e} = \frac{1}{\sqrt{N/2 - 1}} \sum_{i \in \text{white}} \ket{i}. \\
%	&\ket{c} = \ket{\text{yellow}},
%\end{align*}
\begin{align*}
	\ket{a} &= \ket{\text{red}} \\
	\ket{b} &= \frac{1}{\sqrt{N/2 - 2}} \sum_{i \in \text{blue}} \ket{i} \\
	\ket{c} &= \ket{\text{yellow}} \\
	\ket{d} &= \ket{\text{green}} \\
	\ket{e} &= \frac{1}{\sqrt{N/2 - 1}} \sum_{i \in \text{white}} \ket{i}.
\end{align*}
Most of the vertices have degree $N/2 - 1$, except for the yellow and green vertices, which have degree $N/2$. But for large $N$, they are asymptotically the same. So we assume that the graph is approximately regular. Then the search Hamiltonian \eqref{eq:H} for large $N$ is
\[ \setlength{\arraycolsep}{1pt} H = -\gamma \! \left( \! \begin{matrix}
	\frac{1}{\gamma} & \sqrt{\frac{N}{2}-2} & 1 & 0 & 0 \\
	\sqrt{\frac{N}{2}-2} & \frac{N}{2} - 3 & \sqrt{\frac{N}{2}-2} & 0 & 0 \\
	1 & \sqrt{\frac{N}{2}-2} & 0 & 1 & 0 \\
	0 & 0 & 1 & 0 & \sqrt{\frac{N}{2}-1} \\
	0 & 0 & 0 & \sqrt{\frac{N}{2}-1} & \frac{N}{2}-2 \\
\end{matrix} \! \right) \! , \setlength{\arraycolsep}{3pt} \]
where the second item in the first row, for example, is from the adjacency matrix, and it is $1/\sqrt{N/2-2}$ to convert between the normalizations of $\ket{a}$ and $\ket{b}$ times the $N/2-2$ blue vertices that connect to the red vertex.

\begin{figure}
\begin{center}
	\includegraphics{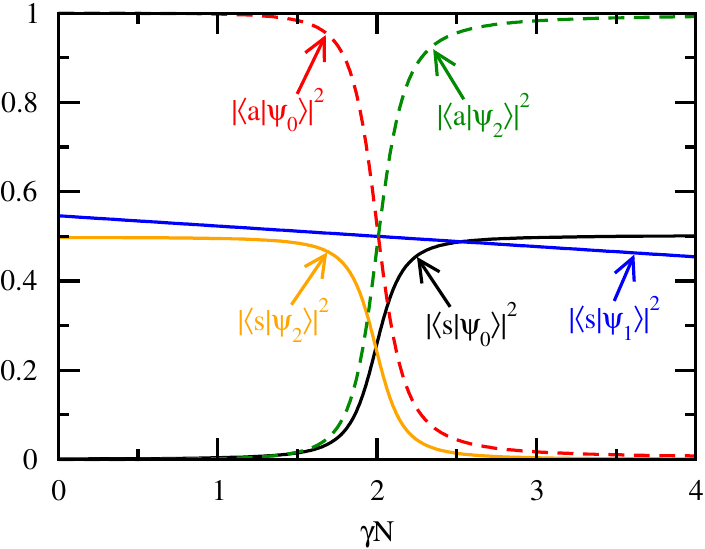}
	\caption{\label{fig:joined_overlap} Squared overlaps of $\ket{s}$ and $\ket{a}$ with eigenstates of $H$ for search on joined complete graphs with 1024 total vertices.}
\end{center}
\end{figure}

\begin{figure}
\begin{center}
	\includegraphics{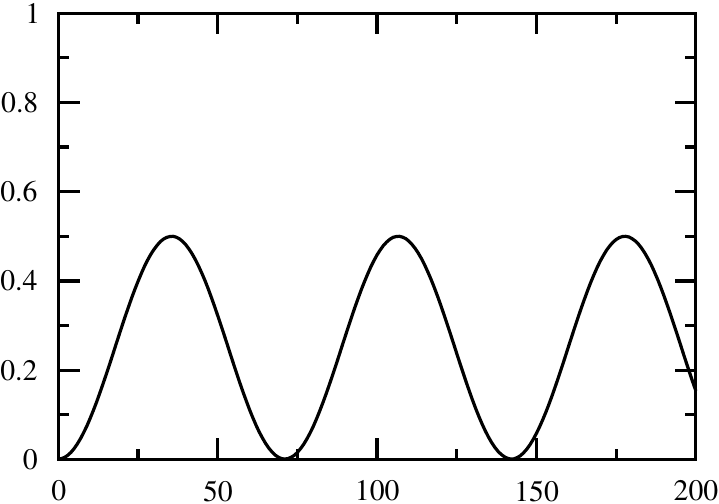}
	\caption{\label{fig:joined_prob_time} Success probability as a function of time for search on joined complete graphs with 1024 total vertices.}
\end{center}
\end{figure}

Fig.~\ref{fig:joined_overlap} shows the squared overlaps of $\ket{s}$ and $\ket{a}$ with the eigenstates of $H$. For large $N$, $\gamma$ takes its critical value of $\gamma_c = 1/(N/2)$, at which half of $\ket{s}$ is proportional to $\ket{\psi_0} + \ket{\psi_2}$ (with the other half in $\ket{\psi_1}$) and $\ket{a} \propto \ket{\psi_0} - \ket{\psi_2}$ with an energy gap of $E_2 - E_0 = 2/\sqrt{N/2}$ \footnotemark[1]. Comparing this to \eqref{eq:complete_sa}, this is the same as searching on a complete graph with $N/2$ vertices and total probability $1/2$, which proves that the success probability reaches $1/2$ in time $\pi \sqrt{N/2} / 2$. This can be seen in Fig.~\ref{fig:joined_prob_time}. 

%-------------------------------------------------------------------------------
% Section.
%-------------------------------------------------------------------------------

\textit{Simplex of Complete Graphs.}---For the second example, we replace each of the $M+1$ vertices of an $M$-simplex with a complete graph of $M$ vertices. An example with $M = 5$ is shown in Fig.~\ref{fig:simplex}; the marked vertex is colored red, and identically evolving vertices are the same color. Note the vertices are homogeneous (\textit{i.e.}, the graph is vertex transitive), and there are $N = M(M+1)$ total vertices. More formally, this is a first-order truncated $M$-simplex lattice, which has been studied in various statistical mechanics models \cite{NF1975, Dhar1977}; the infinite-order recursive construction has also been studied using classical random walks \cite{KG2003}.

From Fig.~\ref{fig:simplex}, the system evolves in a 7-dimensional subspace, independent of $M$. Grouping identically-evolving vertices, we get an orthonormal basis for this subspace:
\begin{align*}
	\ket{a} &= \ket{\text{red}} \\
	\ket{b} &= \frac{1}{\sqrt{M-1}} \sum_{i \in \text{blue}} \ket{i} \\
	\ket{c} &= \ket{\text{yellow}} \\
	\ket{d} &= \frac{1}{\sqrt{M-1}} \sum_{i \in \text{magenta}} \ket{i} \\
	\ket{e} &= \frac{1}{\sqrt{M-1}} \sum_{i \in \text{green}} \ket{i} \\
	\ket{f} &= \frac{1}{\sqrt{M-1}} \sum_{i \in \text{brown}} \ket{i} \\
	\ket{g} &= \frac{1}{\sqrt{(M-1)(M-2)}} \sum_{i \in \text{white}} \ket{i}.
\end{align*}
Then the Hamiltonian \eqref{eq:H} in this subspace is
\[ \setlength{\arraycolsep}{1pt} H = -\gamma \! \left( \! \begin{matrix}
	\frac{1}{\gamma} & \sqrt{M_1} & 1 & 0 & 0 & 0 & 0 \\
	\sqrt{M_1} & M_2 & 0 & 0 & 1 & 0 & 0 \\
	1 & 0 & 0 & \sqrt{M_1} & 0 & 0 & 0 \\
	0 & 0 & \sqrt{M_1} & M_2 & 0 & 1 & 0 \\
	0 & 1 & 0 & 0 & 0 & 1 & \sqrt{M_2} \\
	0 & 0 & 0 & 1 & 1 & 0 & \sqrt{M_2} \\
	0 & 0 & 0 & 0 & \sqrt{M_2} & \sqrt{M_2} & M_2 \\
\end{matrix} \! \right) \! , \setlength{\arraycolsep}{3pt} \]
where $M_1 = M - 1$ and $M_2 = M-2$. The last item in the sixth row, for example, is from the adjacency matrix, and it is $\sqrt{M-1}/\sqrt{(M-1)(M-2)}$ to convert between the normalizations of $\ket{g}$ and $\ket{f}$ times the $M-2$ white vertices that connect to a brown vertex.

\begin{figure}
\begin{center}
	\includegraphics{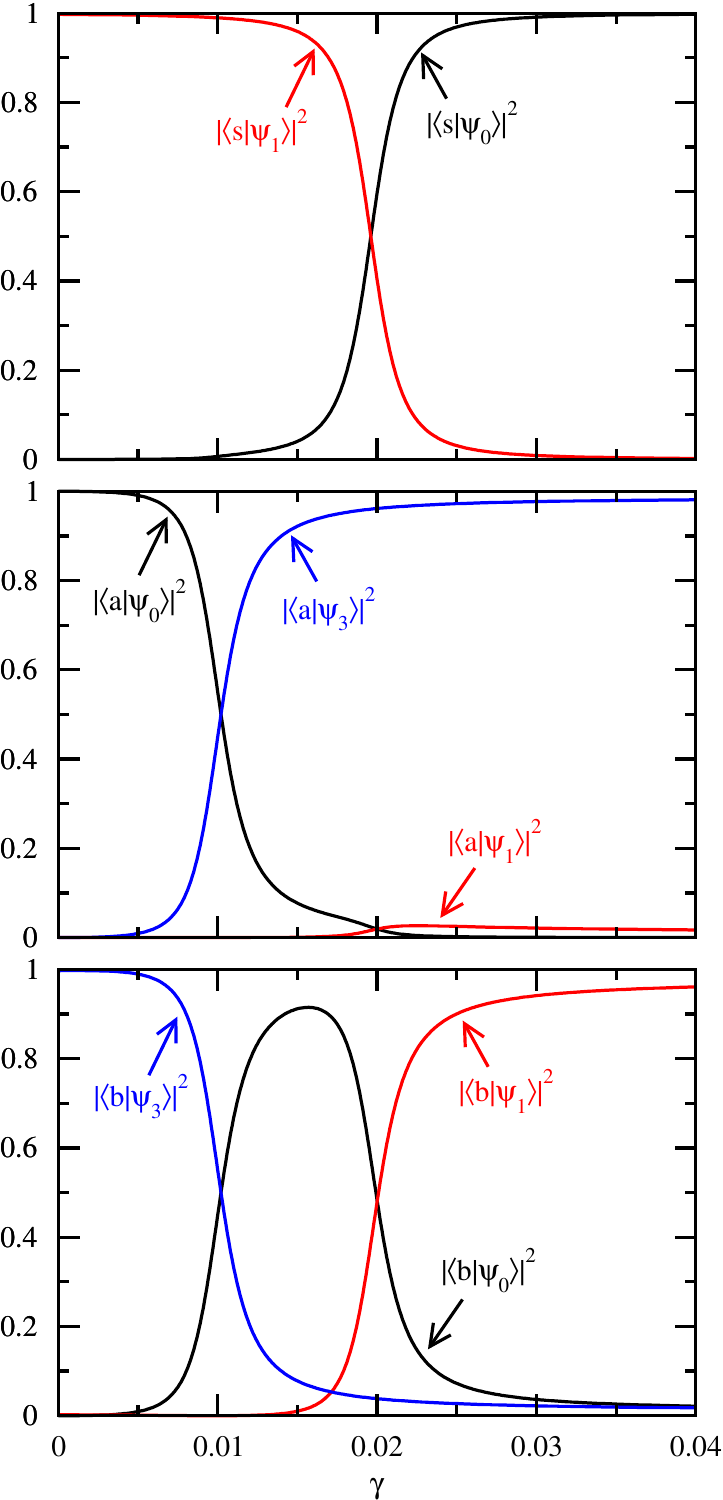}
	\caption{\label{fig:simplex_overlap} Squared overlaps of $\ket{s}$, $\ket{a}$, and $\ket{b}$ with eigenstates of $H$ for search on a simplex of complete graphs with $M = 100$.}
\end{center}
\end{figure}

\begin{figure}
\begin{center}
	\includegraphics{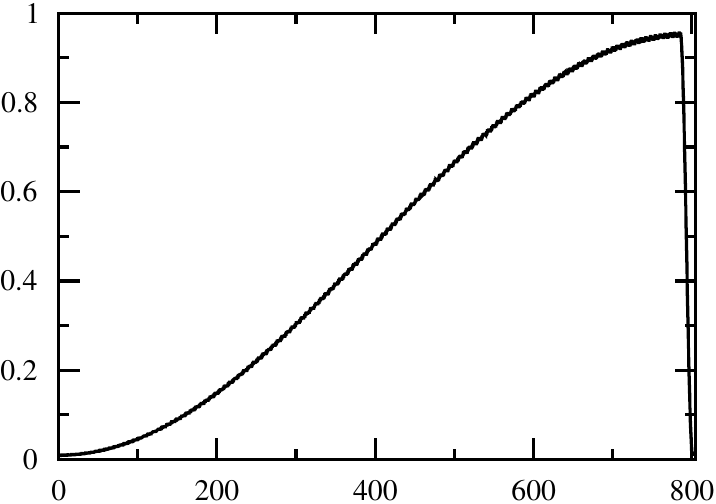}
	\caption{\label{fig:simplex_prob_time_b} Probability at $\ket{b}$ as a function of time for search on a simplex of complete graphs with $M = 100$. Probability accumulates during the first stage of the algorithm from $t = 0$ to $\pi 100^{3/2}/4 \approx 785.40$, and then it quickly leaves during the second stage which takes $\pi\sqrt{100}/2 \approx 15.71$ time.}
\end{center}
\end{figure}

Fig.~\ref{fig:simplex_overlap} shows the squared overlaps of $\ket{s}$, $\ket{a}$, and $\ket{b}$ with the eigenstates of $H$, and it reveals a novel two-stage algorithm. First we let $\gamma$ equal $\gamma_{c1} = 2/M$, which is $0.02$ in Fig.~\ref{fig:simplex_overlap}, because away from this critical value, the initial equal superposition $\ket{s}$ would approximately be an eigenstate of $H$ for large $N$, and then the system would approximately evolve only by phase multiplication. At this critical $\gamma$, roughly $\ket{s} \propto \ket{\psi_0} + \ket{\psi_1}$ and $\ket{b} \propto \ket{\psi_0} - \ket{\psi_1}$ with an energy gap of $4/M^{3/2}$ \footnotemark[1]. Comparing this with \eqref{eq:complete_sa}, the system evolves from $\ket{s}$ to $\ket{b}$ in time $\pi M^{3/2} / 4$, as shown in Fig.~\ref{fig:simplex_prob_time_b}.

\begin{figure}
\begin{center}
	\includegraphics{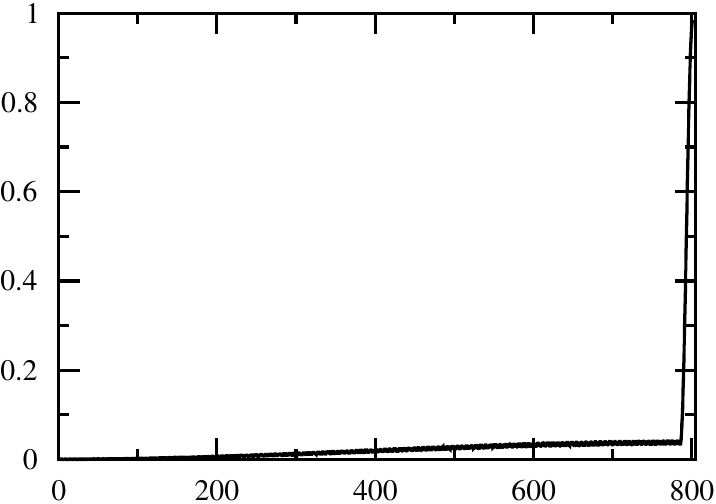}
	\caption{\label{fig:simplex_prob_time_a} Probability at $\ket{a}$ (\textit{i.e.}, the success probability) as a function of time for search on a simplex of complete graphs with $M = 100$. During the second stage of the algorithm starting at $t = \pi 100^{3/2}/4 \approx 785.40$ for a time of $\pi\sqrt{100}/2 \approx 15.75$, the probability quickly accumulates.}
\end{center}
\end{figure}

Now we change $\gamma$ so it equals $\gamma_{c2} = 1/M$, which is $0.01$ in Fig.~\ref{fig:simplex_overlap}. While changing $\gamma$ continously appears in our nonlinear (quantum) search algorithms \cite{MeyerWong2013, MeyerWong2014}, such a discrete change is new. Then roughly $\ket{b} \propto \ket{\psi_0} + \ket{\psi_3}$ and $\ket{a} \propto \ket{\psi_0} - \ket{\psi_3}$ with an energy gap of $E_3 - E_0 = 2/\sqrt{M}$ \footnotemark[1]. Comparing this with \eqref{eq:complete_sa}, probability moves from $\ket{b}$ to $\ket{a}$ in time $\pi \sqrt{M} / 2$, as shown in Figs.~\ref{fig:simplex_prob_time_b} and \ref{fig:simplex_prob_time_a} with $M = 100$, where the sudden dip and spike occurs when switching from the first to the second stage of the algorithm at $\pi 100^{3/2}/4 \approx 785.40$.

Together, the total runtime of this two-stage algorithm is $\pi M^{3/2} / 4 + \pi \sqrt{M} / 2 = \Theta(N^{3/4})$, which is slower than the $\Theta(\sqrt{N})$, $\Theta(\sqrt{N})$, and $\Theta(\sqrt{N} \log^{3/2} N)$ runtimes that vertex/edge, algebraic, and normalized algebraic connectivites would suggest by comparison to the (strongly regular) Latin square graph, hypercube, and 4-dimensional cubic lattice, respectively.

These examples demonstrate that there is not a tight relationship between any of these measures of connectivity and the runtime of quantum random walk search algorithms, disproving the intuition that quantum search is fast as a consequence of high connectivity.

\begin{acknowledgments}
	Thanks to Andrew Childs for confirming our result for search on a 2D square lattice \cite{Childs2012}, and thanks to Andris Ambainis for suggesting search on two complete graphs joined by a single edge. This work was partially supported by the Defense Advanced Research Projects Agency as part of the Quantum Entanglement Science and Technology program under grant N66001-09-1-2025, the Air Force Office of Scientific Research as part of the Transformational Computing in Aerospace Science and Engineering Initiative under grant FA9550-12-1-0046, and the Achievement Awards for College Scientists Foundation.
\end{acknowledgments}

%-------------------------------------------------------------------------------
% References.
%-------------------------------------------------------------------------------

\bibliography{refs}

%-------------------------------------------------------------------------------
% Supplemental Materials.
% Uncomment this for the arXiv version.
%-------------------------------------------------------------------------------

\makeatletter
\close@column@grid
\cleardoublepage
\makeatother
\includepdf[pages={{},1-10}]{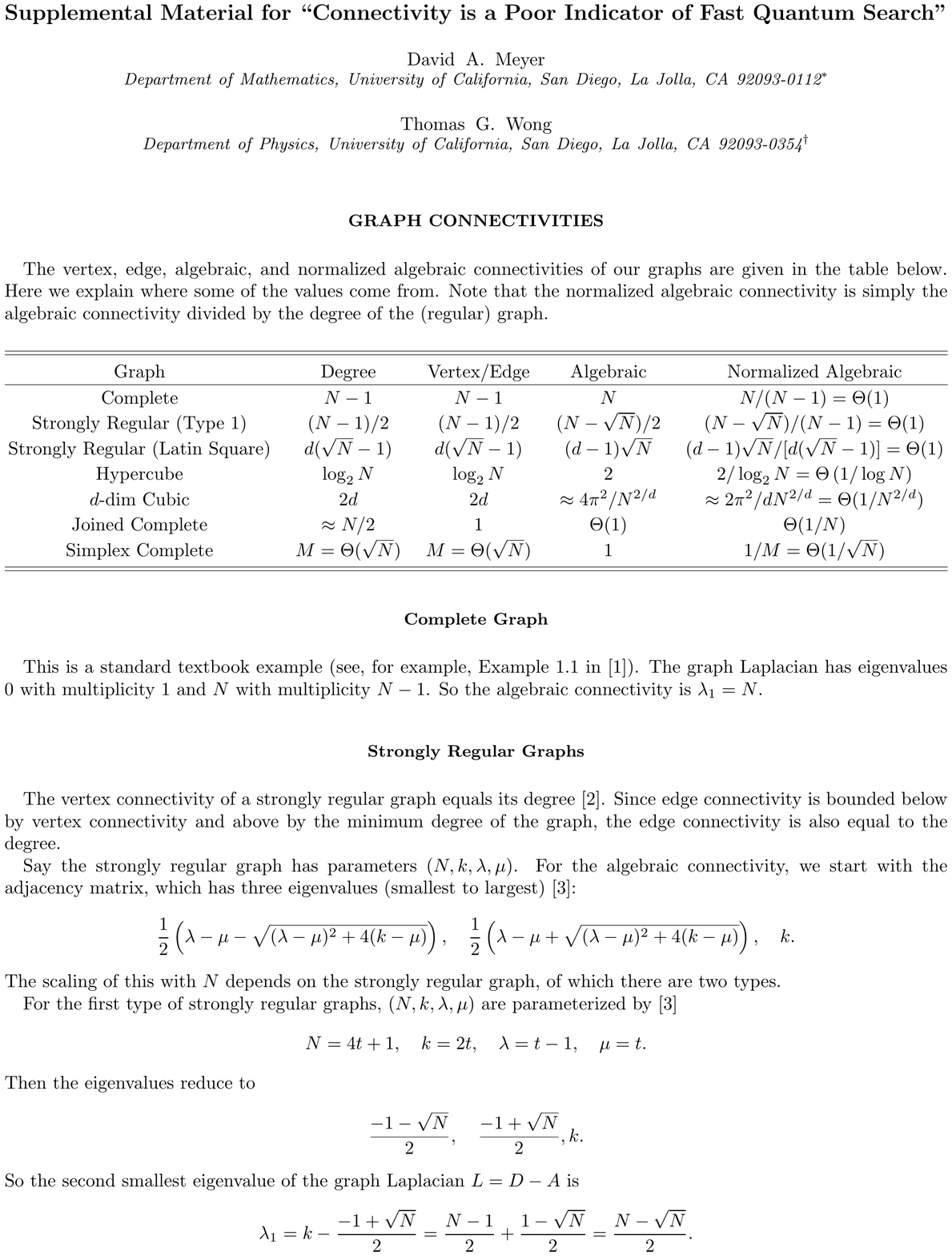}\AtBeginShipout\AtBeginShipoutDiscard

\end{document}

%-------------------------------------------------------------------------------
% Shortcut Commands
%-------------------------------------------------------------------------------

\newcommand{\comb}[2]{{\begin{pmatrix} #1 \\ #2 \end{pmatrix}}}
\newcommand{\braket}[2]{{\left\langle #1 \middle| #2 \right\rangle}}
\newcommand{\bra}[1]{{\left\langle #1 \right|}}
\newcommand{\ket}[1]{{\left| #1 \right\rangle}}
\newcommand{\ketbra}[2]{{\left| #1 \middle\rangle \middle \langle #2 \right|}}

%-------------------------------------------------------------------------------
% Front Matter
%-------------------------------------------------------------------------------

\title{Supplemental Material for ``Connectivity is a Poor Indicator of Fast Quantum Search''}

\author{David A. Meyer}
	\affiliation{Department of Mathematics, University of California, San Diego, La Jolla, CA 92093-0112}
	\email{dmeyer@math.ucsd.edu}
\author{Thomas G. Wong}
	\affiliation{Department of Physics, University of California, San Diego, La Jolla, CA 92093-0354}
	\altaffiliation{Currently at the University of Latvia}
	\email{twong@lu.lv}

\maketitle

%-------------------------------------------------------------------------------
% Main Matter
%-------------------------------------------------------------------------------

\section{Graph Connectivities}

The vertex, edge, algebraic, and normalized algebraic connectivities of our graphs are given in the table below. Here we explain where some of the values come from. Note that the normalized algebraic connectivity is simply the algebraic connectivity divided by the degree of the (regular) graph.

\vspace{.2in}

\begin{ruledtabular}
\begin{tabular}{ccccc}
	Graph & Degree & Vertex/Edge & Algebraic & Normalized Algebraic \\
	\colrule
	Complete & $N-1$ & $N-1$ & $N$ & $N/(N-1) = \Theta(1)$ \\
	Strongly Regular (Type 1) & $(N-1)/2$ & $(N-1)/2$ & $(N - \sqrt{N})/2$ & $(N - \sqrt{N})/(N - 1) = \Theta(1)$ \\
	Strongly Regular (Latin Square) & $d(\sqrt{N}-1)$ & $d(\sqrt{N}-1)$ & $(d-1)\sqrt{N}$ & $(d-1)\sqrt{N} / [d(\sqrt{N}-1)] = \Theta(1)$ \\
	Hypercube & $\log_2 N$ & $\log_2 N$ & $2$ & $2 / \log_2 N = \Theta \left( 1 / \log N \right)$ \\
	$d$-dim Cubic & $2d$ & $2d$ & $\approx 4 \pi^2 / N^{2/d}$ & $\approx 2 \pi^2 / d N^{2/d} = \Theta( 1 / N^{2/d} )$ \\
	Joined Complete & $\approx N/2$ & $1$ & $\Theta(1)$ & $\Theta(1/N)$ \\
	Simplex Complete & $M = \Theta(\sqrt{N})$ & $M = \Theta(\sqrt{N})$ & $1$ & $1/M = \Theta(1/\sqrt{N})$
\end{tabular}
\end{ruledtabular}

\subsection{Complete Graph}

	This is a standard textbook example (see, for example, Example 1.1 in \cite{Chung1997}). The graph Laplacian has eigenvalues $0$ with multiplicity 1 and $N$ with multiplicity $N - 1$. So the algebraic connectivity is $\lambda_1 = N$.

\subsection{Strongly Regular Graphs}

	The vertex connectivity of a strongly regular graph equals its degree \cite{BM1985}. Since edge connectivity is bounded below by vertex connectivity and above by the minimum degree of the graph, the edge connectivity is also equal to the degree.

	Say the strongly regular graph has parameters $(N,k,\lambda,\mu)$. For the algebraic connectivity, we start with the adjacency matrix, which has three eigenvalues (smallest to largest) \cite{Cameron1991}:
	\[ \frac{1}{2} \left( \lambda - \mu - \sqrt{(\lambda - \mu)^2 + 4(k - \mu)} \right), \quad \frac{1}{2} \left( \lambda - \mu + \sqrt{(\lambda - \mu)^2 + 4(k - \mu)} \right), \quad k. \]
	The scaling of this with $N$ depends on the strongly regular graph, of which there are two types.

	For the first type of strongly regular graphs, $(N,k,\lambda,\mu)$ are parameterized by \cite{Cameron1991}
	\[ N = 4t+1, \quad k = 2t, \quad \lambda = t - 1, \quad \mu = t. \]
	Then the eigenvalues reduce to 
	\[ \frac{-1 - \sqrt{N}}{2}, \quad \frac{-1 + \sqrt{N}}{2}, k .\]
	So the second smallest eigenvalue of the graph Laplacian $L = D - A$ is
	\[ \lambda_1 = k - \frac{-1 + \sqrt{N}}{2} = \frac{N-1}{2} + \frac{1 - \sqrt{N}}{2} = \frac{N - \sqrt{N}}{2}. \]

	Not all strongly regular graphs of the second type are known, but certain parameter families are. Here, we give the example of Latin square graphs, which have
	\[ N = t^2, \, k = d(t-1), \, \lambda = d^2 - 3d + t, \, \text{and} \, \mu = d(d-1). \]
	With these parameters, the second smallest eigenvalue of the graph Laplacian is
	\[ \lambda_1 = (d - 1) \sqrt{N}. \]

\subsection{Hypercube}

	This is a standard textbook example (see, for example, Example 1.6 in \cite{Chung1997}). The graph Laplacian has eigenvalues $2k$ for $k = 0, 1, \dots, n$ with multiplicities ``$n$ choose $k$.'' So the algebraic connectivity is $\lambda_1 = 2$.

\subsection{Cubic}

	This is a standard textbook problem (see, for example, Section 4.3 of \cite{Sakurai1993} or Section 5.3.2 of \cite{Griffiths2005}). The eigenvalues of the graph Laplacian for a $d$-dimensional cubic lattice are
	\[ \lambda(\vec{k}) = 2 \left( d - \sum{j = 1}^d \cos(k_j) \right), \]
	where
	\[ k_j = \frac{2 \pi m_j}{N^{1/d}} \]
	and
	\[ m_j = \begin{cases}
		0, \pm 1, \pm 2, \dots, \pm \frac{1}{2} \left( N^{1/d} - 1 \right) & N^{1/d} \text{ odd} \\
			0, \pm 1, \pm 2, \dots, \pm \frac{1}{2} \left( N^{1/d} - 2 \right), +\frac{1}{2} N^{1/d} & N^{1/d} \text{ even} \\
	\end{cases}. \]
	This takes its smallest value of $\lambda_0 = 0$ when $\vec{k} = 0$. It takes its next smallest value of $\lambda_1 = 2 \left( 1 - \cos \frac{2\pi}{N^{1/d}} \right)$ when one of the $k_j's$ is $1$ and the rest are $0$. For large $N$, this can be Taylor expanded to yield $(2 \pi / N^{1/d})^2$.

\subsection{Joined Complete Graphs}

	The adjacency matrix in the 5-dimensional subspace is
	\[ A = \begin{pmatrix}
		0 & \sqrt{N/2-2} & 1 & 0 & 0 \\
		\sqrt{N/2-2} & N/2 - 3 & \sqrt{N/2-2} & 0 & 0 \\
		1 & \sqrt{N/2-2} & 0 & 1 & 0 \\
		0 & 0 & 1 & 0 & \sqrt{N/2-1} \\
		0 & 0 & 0 & \sqrt{N/2-1} & N/2-2 \\
	\end{pmatrix}, \]
	and it has eigenvalues
	\begin{gather*}
		\frac{1}{4} \left( N - 6 - \sqrt{(N+6)(N-2)} \right) \approx -2 + \frac{2}{N} \\
		-1 \\
		\frac{1}{4} \left( N - 2 - \sqrt{N^2 - 4N + 20} \right) \approx -\frac{2}{N} - \frac{4}{N^2} \\
		\frac{1}{4} \left( N - 6 + \sqrt{(N+6)(N-2)} \right) \approx \frac{N}{2} - 1 - \frac{2}{N} \\
		\frac{1}{4} \left( N - 2 + \sqrt{N^2 - 4N + 20} \right) \approx \frac{N}{2} - 1 + \frac{2}{N}
	\end{gather*}
	Since $L = D - A$, the largest eigenvalue of $A$ corresponds to the smallest eigenvalue of $L$. Assuming the graph is approximately regular with degree $N/2 - 1$, the two smallest eigenvalues of $L$ are $-2/N$ and $2/N$. But this can not be right---$L$ is positive definite. The discrepancy is from our assumption that the graph is regular when it is not; the yellow and green vertices have degree $N/2$, not $N/2 - 1$, so we have made an error that is $\Theta(1)$. So the algebraic connectivity is $\lambda_1 = \Theta(1)$.

\subsection{Simplex of Complete Graphs}

	The graph Laplacian in the 7-dimensional subspace is
	\[ L = \begin{pmatrix}
		M & -\sqrt{M-1} & -1 & 0 & 0 & 0 & 0 \\
		-\sqrt{M-1} & 2 & 0 & 0 & -1 & 0 & 0 \\
		-1 & 0 & M & -\sqrt{M-1} & 0 & 0 & 0 \\
		0 & 0 & -\sqrt{M-1} & 2 & 0 & -1 & 0 \\
		0 & -1 & 0 & 0 & M & -1 & -\sqrt{M-2} \\
		0 & 0 & 0 & -1 & -1 & M & -\sqrt{M-2} \\
		0 & 0 & 0 & 0 & -\sqrt{M-2} & -\sqrt{M-2} & 2 \\
	\end{pmatrix}. \]
	Its eigenvalues are
	\[ 0, 1, 1, M, M+1, M+1, M+2. \]
	So the algebraic connectivity is $\lambda_1 = 1$.

%-------------------------------------------------------------------------------
% Section.
%-------------------------------------------------------------------------------

\clearpage

\section{Perturbative Solution for Joined Complete Graphs}

\begin{center}
	\includegraphics{joined.pdf}
\end{center}

For large $N$, the Hamiltonian is
\[ H = -\gamma \begin{pmatrix}
	\frac{1}{\gamma} & \sqrt{\frac{N}{2}} & 1 & 0 & 0 \\
	\sqrt{\frac{N}{2}} & \frac{N}{2} & \sqrt{\frac{N}{2}} & 0 & 0 \\
	1 & \sqrt{\frac{N}{2}} & 0 & 1 & 0 \\
	0 & 0 & 1 & 0 & \sqrt{\frac{N}{2}} \\
	0 & 0 & 0 & \sqrt{\frac{N}{2}} & \frac{N}{2} \\
\end{pmatrix}. \]
To do perturbation theory, we break the Hamiltonian into leading- and higher-order terms. We get
\[ H^{(0)} = -\gamma \begin{pmatrix}
	\frac{1}{\gamma} & 0 & 0 & 0 & 0 \\
	0 & \frac{N}{2} & 0 & 0 & 0 \\
	0 & 0 & 0 & 0 & 0 \\
	0 & 0 & 0 & 0 & 0 \\
	0 & 0 & 0 & 0 & \frac{N}{2} \\
\end{pmatrix}, \quad H^{(1)} = -\gamma \sqrt{\frac{N}{2}} \begin{pmatrix}
	0 & 1 & 0 & 0 & 0 \\
	1 & 0 & 1 & 0 & 0 \\
	0 & 1 & 0 & 0 & 0 \\
	0 & 0 & 0 & 0 & 1 \\
	0 & 0 & 0 & 1 & 0 \\
\end{pmatrix}, \quad H^{(2)} = -\gamma \begin{pmatrix}
	0 & 0 & 1 & 0 & 0 \\
	0 & 0 & 0 & 0 & 0 \\
	1 & 0 & 0 & 1 & 0 \\
	0 & 0 & 1 & 0 & 0 \\
	0 & 0 & 0 & 0 & 0 \\
\end{pmatrix}.\]
We are only doing this to first order, so we can ignore $H^{(2)}$.

It is clear that $\ket{a}$, $\ket{b}$, and $\ket{e}$ are eigenvectors of $H^{(0)}$ with corresponding eigenvalues $-1$, $-\gamma \frac{N}{2}$, and $-\gamma \frac{N}{2}$. For these to be (triply) degenerate, we need
\[ \gamma_c = \frac{2}{N}. \]
The perturbation lifts the degeneracy, and the eigenvectors of the perturbed system will be linear combinations of $\ket{a}$, $\ket{b}$, and $\ket{e}$:
\[ \ket{\psi} = \alpha_a \ket{a} + \alpha_b \ket{b} + \alpha_e \ket{e}. \]
The coefficients can be found by solving
\[ \begin{pmatrix}
	H_{aa} & H_{ab} & H_{ae} \\
	H_{ba} & H_{bb} & H_{be} \\
	H_{ea} & H_{eb} & H_{ee} \\
\end{pmatrix} \begin{pmatrix}
	\alpha_a \\
	\alpha_b \\
	\alpha_e \\
\end{pmatrix} = E \begin{pmatrix}
	\alpha_a \\
	\alpha_b \\
	\alpha_e \\
\end{pmatrix}, \]
where $H_{ab} = \bra{a} H^{(0)} + H^{(1)} \ket{b}$, etc. Evaluating these matrix components with $\gamma = \gamma_c$, we get
\[ \begin{pmatrix}
	-1 & -\sqrt{\frac{2}{N}} & 0 \\
	-\sqrt{\frac{2}{N}} & -1 & 0 \\
	0 & 0 & -1 \\
\end{pmatrix} \begin{pmatrix}
	\alpha_a \\
	\alpha_b \\
	\alpha_e \\
\end{pmatrix} = E \begin{pmatrix}
	\alpha_a \\
	\alpha_b \\
	\alpha_e \\
\end{pmatrix}. \]
The solution to this yields the ground, first excited, and second excited states and their corresponding eigenvalues:
\begin{gather*}
	\ket{\psi_0} = \frac{1}{\sqrt{2}} \begin{pmatrix} 1 \\ 1 \\ 0 \end{pmatrix} = \frac{1}{\sqrt{2}} (\ket{a} + \ket{b}), \quad E_0 = -1 - \sqrt{\frac{2}{N}} \\
	\ket{\psi_1} = \begin{pmatrix} 0 \\ 0 \\ 1 \end{pmatrix} = \ket{e}, \quad E_1 = -1 \\
	\ket{\psi_2} = \frac{1}{\sqrt{2}} \begin{pmatrix} -1 \\ 1 \\ 0 \end{pmatrix} = \frac{1}{\sqrt{2}} (-\ket{a} + \ket{b}), \quad E_2 = -1 + \sqrt{\frac{2}{N}}.
\end{gather*}

Using these eigenstates and eigenvalues, we can find the evolution of the system, runtime, and success probability. Note that the system starts in the equal superposition of all vertices, which in the 5-dimensional subspace is
\[ \ket{s} = \frac{1}{\sqrt{N}} \ket{a} + \frac{\sqrt{N/2 - 2}}{\sqrt{N}} \ket{b} + \frac{1}{\sqrt{N}} \ket{c} + \frac{1}{\sqrt{N}} \ket{d} + \frac{\sqrt{N/2 - 1}}{\sqrt{N}} \ket{e}. \]
To leading order, this is dominated by $\ket{b}$ and $\ket{e}$:
\[ \ket{s} \approx \frac{1}{\sqrt{2}} \left( \ket{b} + \ket{e} \right). \]
But this is just the sum of the first three eigenstates:
\[ \ket{s} \approx \frac{1}{\sqrt{2}} \left( \frac{1}{\sqrt{2}} \left( \ket{\psi_0} + \ket{\psi_2} \right) + \ket{\psi_1} \right). \]
So the system approximately evolves in its three lowest energy eigenstates. So the leading- and first-order evolution is
\[ \ket{\psi(t)} = e^{-i (H^{(0)} + H^{(1)}) t} \ket{s}. \]
Then the success amplitude is
\[ \braket{a}{\psi(t)} \approx \frac{1}{\sqrt{2}} \left( \frac{1}{\sqrt{2}} \left( e^{-iE_0t} \braket{a}{\psi_0} + e^{-iE_2t} \braket{a}{\psi_2} \right) + e^{-iE_1t} \underbrace{\braket{a}{\psi_1}}_0 \right). \]
Note that
\[ \braket{a}{\psi_{0,2}} = \frac{1}{\sqrt{2}}. \]
So
\[ \braket{a}{\psi(t)} \approx \frac{1}{\sqrt{2}} \frac{1}{2} e^{it} \left( e^{i \sqrt{2/N} t} - e^{-i \sqrt{2/N} t} \right) = \frac{1}{\sqrt{2}} e^{it} i \sin \left( \sqrt{\frac{2}{N}} t \right). \]
Then the success probability is
\[ |\braket{a}{\psi(t)}|^2 \approx \frac{1}{2} \sin^2 \left( \sqrt{\frac{2}{N}} t \right), \]
which reaches a max value of $1/2$ at time
\[ \frac{\pi}{2} \sqrt{\frac{N}{2}}. \]

%-------------------------------------------------------------------------------
% Section.
%-------------------------------------------------------------------------------

\clearpage

\section{Perturbative Solution for Simplex of Complete Graphs}

\begin{center}
	\includegraphics{simplex.pdf}
\end{center}

The search Hamiltonian is
\[ H = -\gamma \begin{pmatrix}
	\frac{1}{\gamma} & \sqrt{M-1} & 1 & 0 & 0 & 0 & 0 \\
	\sqrt{M-1} & M-2 & 0 & 0 & 1 & 0 & 0 \\
	1 & 0 & 0 & \sqrt{M-1} & 0 & 0 & 0 \\
	0 & 0 & \sqrt{M-1} & M-2 & 0 & 1 & 0 \\
	0 & 1 & 0 & 0 & 0 & 1 & \sqrt{M-2} \\
	0 & 0 & 0 & 1 & 1 & 0 & \sqrt{M-2} \\
	0 & 0 & 0 & 0 & \sqrt{M-2} & \sqrt{M-2} & M-2 \\
\end{pmatrix}. \]
Let us derive the evolution of each stage of the algorithm.

\subsection{First Stage}

	For the first stage of the algorithm, we choose the unperturbed Hamiltonian to be
	\[ H^{(0)} = -\gamma \begin{pmatrix}
		\frac{1}{\gamma} & \sqrt{M} & 0 & 0 & 0 & 0 & 0 \\
		\sqrt{M} & M & 0 & 0 & 0 & 0 & 0 \\
		0 & 0 & 0 & \sqrt{M} & 0 & 0 & 0 \\
		0 & 0 & \sqrt{M} & M & 0 & 0 & 0 \\
		0 & 0 & 0 & 0 & 0 & 0 & \sqrt{M} \\
		0 & 0 & 0 & 0 & 0 & 0 & \sqrt{M} \\
		0 & 0 & 0 & 0 & \sqrt{M} & \sqrt{M} & M \\
	\end{pmatrix}, \]
	and the perturbation $H^{(1)}$ is the three $-2$'s on the diagonal of $H$, eight $1$'s on the off-diagonal, and zeros elsewhere. Then $H^{(0)}$ has eigenvalues and (unnormalized) eigenvectors
	\begin{gather*}
		0, \quad \{0,0,0,0,-1,1,0\} \\
		-\frac{1}{2} \sqrt{M} \left(\sqrt{M}+\sqrt{M+4}\right) \gamma, \quad \left\{0,0,\frac{1}{2} \left(-\sqrt{M}+\sqrt{M+4}\right),1,0,0,0\right\} \\
		-\frac{1}{2} \sqrt{M} \left(\sqrt{M}-\sqrt{M+4}\right) \gamma, \quad \left\{0,0,\frac{1}{2} \left(-\sqrt{M}-\sqrt{M+4}\right),1,0,0,0\right\}
	\end{gather*}
	\begin{gather*}
		-\frac{1}{2} \sqrt{M} \left(\sqrt{M}+\sqrt{M+8}\right) \gamma, \quad \left\{0,0,0,0,\frac{2}{\sqrt{M}+\sqrt{M+8}},\frac{2}{\sqrt{M}+\sqrt{M+8}},1\right\} \\
		-\frac{1}{2} \sqrt{M} \left(\sqrt{M}-\sqrt{M+8}\right) \gamma, \quad \left\{0,0,0,0,\frac{2}{\sqrt{M}-\sqrt{M+8}},\frac{2}{\sqrt{M}-\sqrt{M+8}},1\right\} \\
		-\frac{1}{2} \left(1+M \gamma +\sqrt{1-2 M \gamma +4 M \gamma ^2+M^2 \gamma ^2}\right), \quad \left\{\frac{1-M \gamma +\sqrt{1-2 M \gamma +4 M \gamma ^2+M^2 \gamma ^2}}{2 \sqrt{M} \gamma },1,0,0,0,0,0\right\} \\
		-\frac{1}{2} \left(1+M \gamma -\sqrt{1-2 M \gamma +4 M \gamma ^2+M^2 \gamma ^2}\right), \quad \left\{\frac{1-M \gamma -\sqrt{1-2 M \gamma +4 M \gamma ^2+M^2 \gamma ^2}}{2 \sqrt{M} \gamma },1,0,0,0,0,0\right\}
	\end{gather*}
	We want to choose $\gamma$ positive and real so that an eigenstate with a large projection on $\ket{g}$ is degenerate with an eigenstate with a large projection on $\ket{b}$. When
	\[ \gamma_{c1} = \frac{-M + \sqrt{M}\sqrt{8+M}}{2M} \approx \frac{2}{M}, \]
	the fourth and sixth eigenvalues are degenerate and both equal $-2$. Let us call the corresponding eigenvectors (\textit{i.e.}, the forth and sixth eigenstates above) $u$ and $v$ and their normalized versions $\ket{u}$ and $\ket{v}$.
	
	The perturbation lifts the degeneracy such that the eigenstates become
	\[ \ket{\psi} = \alpha_u \ket{u} + \alpha_v \ket{v}. \]
	The coefficients can be found by solving
	\[ \begin{pmatrix}
		H_{uu} & H_{uv} \\
		H_{vu} & H_{vv} \\
	\end{pmatrix} \begin{pmatrix}
		\alpha_u \\
		\alpha_v \\
	\end{pmatrix} = E \begin{pmatrix}
		\alpha_u \\
		\alpha_v \\
	\end{pmatrix}, \]
	where $H_{uv} = \bra{u} H^{(0)} + H^{(1)} \ket{v}$, etc. Using the ``full'' $\ket{u}$ and $\ket{v}$ when computing $H_{uv}$ and letting $\gamma = \gamma_{c1}$, the terms $O(1/M^3)$ are
	\[ \begin{pmatrix}
		-2 + \frac{4}{M} - \frac{20}{M^2} + \frac{128}{M^3} + O\left(\frac{1}{M^4}\right) & -\frac{2}{M^{3/2}} + \frac{14}{M^{5/2}} + O\left(\frac{1}{M^{7/2}}\right) \\
		-\frac{2}{M^{3/2}} + \frac{14}{M^{5/2}} + O\left(\frac{1}{M^{7/2}}\right) & -2 + \frac{4}{M} - \frac{24}{M^2} + \frac{192}{M^3} + O\left(\frac{1}{M^4}\right) \\
	\end{pmatrix} \begin{pmatrix}
		\alpha_u \\
		\alpha_v \\
	\end{pmatrix} = E \begin{pmatrix}
		\alpha_u \\
		\alpha_v \\
	\end{pmatrix}. \]
	Solving this keeping terms $O(1/M^{3/2})$ in the final answer, the (unnormalized) ground and first-excited states and energies are
	\begin{gather*}
		\psi_0 = \left\{ 1 - \frac{1}{\sqrt{M}} + \frac{1}{2M}, 1 \right\} \approx \left\{ 1, 1 \right\}, \quad E_0 = -2 + \frac{4}{M} - \frac{2}{M^{3/2}}, \\
		\psi_1 = \left\{ -1 - \frac{1}{\sqrt{M}} - \frac{1}{2M}, 1 \right\} \approx \left\{ -1, 1 \right\}, \quad E_1 = -2 + \frac{4}{M} + \frac{2}{M^{3/2}}.
	\end{gather*}
	Normalizing the eigenstates,
	\begin{gather*}
		\ket{\psi_0} = \frac{1}{\sqrt{2}} \left( \ket{u} + \ket{v} \right), \quad E_0 = -2 + \frac{4}{M} - \frac{2}{M^{3/2}}, \\
		\ket{\psi_0} = \frac{1}{\sqrt{2}} \left( -\ket{u} + \ket{v} \right), \quad E_1 = -2 + \frac{4}{M} + \frac{2}{M^{3/2}}.
	\end{gather*}
	Note that up to terms of $O(1/M)$,
	\begin{gather*}
		\ket{u} \approx \sqrt{\frac{M}{M+2}} \left( \frac{1}{\sqrt{M}} \ket{e} + \frac{1}{\sqrt{M}} \ket{f} + \ket{g} \right) \\
		\ket{v} \approx \sqrt{\frac{M}{M+4}} \left( \frac{2}{\sqrt{M}} \ket{a} + \ket{b} \right).
	\end{gather*}
	So the ground and first excited states are
	\begin{gather*}
		\ket{\psi_0} \approx \frac{1}{\sqrt{2}} \left( \frac{2}{\sqrt{M}} \ket{a} + \ket{b} + \frac{1}{\sqrt{M}} \ket{e} + \frac{1}{\sqrt{M}} \ket{f} + \ket{g} \right) \\
		\ket{\psi_1} \approx \frac{1}{\sqrt{2}} \left( -\frac{2}{\sqrt{M}} \ket{a} - \ket{b} + \frac{1}{\sqrt{M}} \ket{e} + \frac{1}{\sqrt{M}} \ket{f} + \ket{g} \right).
	\end{gather*}

	Now let us show that the system roughly evolves from $\ket{s}$ to $\ket{b}$ for large $N$. Recall that the initial state is the equal superposition state:
	\[ \ket{s} = \frac{1}{\sqrt{N}} \ket{a} + \frac{\sqrt{M-1}}{\sqrt{N}} \ket{b} + \frac{1}{\sqrt{N}} \ket{c} + \frac{\sqrt{M-1}}{\sqrt{N}} \ket{d} + \frac{\sqrt{M-1}}{\sqrt{N}} \ket{e} + \frac{\sqrt{M-1}}{\sqrt{N}} \ket{f} + \frac{\sqrt{(M-1)(M-2)}}{\sqrt{N}} \ket{g}. \]
	For large $M$,
	\[ \ket{s} \approx \ket{g} \approx \frac{1}{\sqrt{2}} \left( \ket{\psi_0} + \ket{\psi_1} \right), \]
	up to terms of order $1/\sqrt{N}$ (since $M = \Theta(\sqrt{N})$). Then the evolution is easy to find:
	\begin{align*}
		\ket{\psi(t)} 
			&= e^{-iHt} \ket{s} \\
			&\approx e^{-iHt} \frac{1}{\sqrt{2}} \left( \ket{\psi_0} + \ket{\psi_1} \right) \\
			&= \frac{1}{\sqrt{2}} \left( e^{-iE_0t} \ket{\psi_0} + e^{-iE_1t} \ket{\psi_1} \right) \\
			&= \frac{1}{\sqrt{2}} e^{i(2+4/M)t} \left( e^{2ti/M^{3/2}} \ket{\psi_0} + e^{-2ti/M^{3/2}} \ket{\psi_1} \right)
	\end{align*}
	Taking the inner product with $\bra{b}$,
	\begin{align*}
		\braket{b}{\psi(t)} 
		&\approx \frac{1}{\sqrt{2}} e^{i(2+4/M)t} \left( e^{2ti/M^{3/2}} \frac{1}{\sqrt{2}} - e^{-2ti/M^{3/2}} \frac{1}{\sqrt{2}} \right) \\
		&= e^{i(2+4/M)t} i \sin \left( \frac{2t}{M^{3/2}} \right).
	\end{align*}
	So the system evolves from $\ket{s}$ to $\ket{b}$ (with a phase) in time
	\[ t_1 = \frac{\pi}{4} M^{3/2}. \]

\subsection{Second Stage}

	For the second stage of the algorithm, we choose the unperturbed Hamiltonian to be
	\[ H^{(0)} = -\gamma \begin{pmatrix}
		\frac{1}{\gamma} & 0 & 0 & 0 & 0 & 0 & 0 \\
		0 & M & 0 & 0 & 0 & 0 & 0 \\
		0 & 0 & 0 & 0 & 0 & 0 & 0 \\
		0 & 0 & 0 & M & 0 & 0 & 0 \\
		0 & 0 & 0 & 0 & 0 & 0 & 0 \\
		0 & 0 & 0 & 0 & 0 & 0 & 0 \\
		0 & 0 & 0 & 0 & 0 & 0 & M \\
	\end{pmatrix}, \]
	and the perturbation $H^{(1)}$ is the $\sqrt{M}$ terms in $H$. When
	\[ \gamma_{c2} = \frac{1}{M}, \]
	then the zeroth-order Hamiltonian is quarticly degenerate. That is, $\ket{a}$, $\ket{b}$, $\ket{d}$, and $\ket{g}$ all have eigenvalue -1.

	The perturbation lifts the degeneracy, and the corresponding eigenvectors will be linear combinations of $\ket{a}$, $\ket{b}$, $\ket{d}$, and $\ket{g}$:
	\[ \ket{\psi} = \alpha_a \ket{a} + \alpha_b \ket{b} + \alpha_d \ket{d} + \alpha_g \ket{g}. \]
	The coefficients can be found by solving
	\[ \begin{pmatrix}
		H_{aa} & H_{ab} & H_{ad} & H_{ag} \\
		H_{ba} & H_{bb} & H_{bd} & H_{bg} \\
		H_{da} & H_{db} & H_{dd} & H_{dg} \\
		H_{ga} & H_{gb} & H_{gd} & H_{gg} \\
	\end{pmatrix} \begin{pmatrix}
		\alpha_a \\
		\alpha_b \\
		\alpha_d \\
		\alpha_g \\
	\end{pmatrix} = E \begin{pmatrix}
		\alpha_a \\
		\alpha_b \\
		\alpha_d \\
		\alpha_g \\
	\end{pmatrix}, \]
	where $H_{ab} = \bra{a} H^{(0)} + H^{(1)} \ket{b}$, etc. Evaluating these matrix components with $\gamma = \gamma_{c2}$, this becomes
	\[ \begin{pmatrix}
		-1 & \frac{-1}{\sqrt{M}} & 0 & 0 \\
		\frac{-1}{\sqrt{M}} & -1 & 0 & 0 \\
		0 & 0 & -1 & 0 \\
		0 & 0 & 0 & -1 \\
	\end{pmatrix} \begin{pmatrix}
		\alpha_a \\
		\alpha_b \\
		\alpha_d \\
		\alpha_g \\
	\end{pmatrix} = E \begin{pmatrix}
		\alpha_a \\
		\alpha_b \\
		\alpha_d \\
		\alpha_g \\
	\end{pmatrix}. \]
	Using Mathematica, this has eigenvectors and corresponding eigenvalues
	\begin{gather*}
		\ket{\psi_0} = \frac{1}{\sqrt{2}} \begin{pmatrix} 1 \\ 1 \\ 0 \\ 0 \end{pmatrix} = \frac{1}{\sqrt{2}} (\ket{a} + \ket{b}), \quad E_0 = -1 - \frac{1}{\sqrt{M}} \\
		\ket{\psi_1} = \begin{pmatrix} 0 \\ 0 \\ 1 \\ 0 \end{pmatrix} = \ket{d}, \quad E_1 = -1 \\
		\ket{\psi_2} = \begin{pmatrix} 0 \\ 0 \\ 0 \\ 1 \end{pmatrix} = \ket{g}, \quad E_2 = -1 \\
		\ket{\psi_3} = \frac{1}{\sqrt{2}} \begin{pmatrix} -1 \\ 1 \\ 0 \\ 0 \end{pmatrix} = \frac{1}{\sqrt{2}} (-\ket{a} + \ket{b}), \quad E_3 = -1 + \frac{1}{\sqrt{M}}.
	\end{gather*}

	Now let us calculate the evolution starting in $\ket{b}$. Since
	\[ \ket{b} = \frac{1}{\sqrt{2}} \left( \ket{\psi_0} + \ket{\psi_3} \right), \]
	it evolves to
	\begin{align*}
		e^{-iHt} \ket{b} 
			&= \frac{1}{\sqrt{2}} \left( e^{-iE_0t} \ket{\psi_0} + e^{-iE_3t} \ket{\psi_3} \right) \\
			&= \frac{1}{\sqrt{2}} e^{it} \left( e^{it/\sqrt{M}} \ket{\psi_0} + e^{-it/\sqrt{M}} \ket{\psi_3} \right).
	\end{align*}
	Taking the inner product with $\bra{a}$ and noting that $\braket{a}{\psi_{0,3}} = \pm 1/\sqrt{2}$, this becomes
	\begin{align*}
		\bra{a} e^{-iHt} \ket{b} 
			&= \frac{1}{2} e^{it} \left( e^{it/\sqrt{M}} - e^{-it/\sqrt{M}} \right) \\
			&= e^{it} i \sin \left( \frac{t}{\sqrt{M}} \right)
	\end{align*}
	So the probability moves from $\ket{b}$ to $\ket{a}$ in time
	\[ t_2 = \frac{\pi \sqrt{M}}{2}. \]

%-------------------------------------------------------------------------------
% References.
%-------------------------------------------------------------------------------

\bibliography{refs}